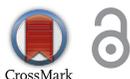

*Corresponding author: Birgitta Dresp-Langley, ICube Lab, CNRS and University of Strasbourg, UMR 7357, Strasbourg, France
E-mail: birgitta.dresp@unistra.fr

# SURGERY | RESEARCH ARTICLE

# Effects of 2D and 3D image views on hand movement trajectories in the surgeon's peri-personal space in a computer controlled simulator environment

Anil Ufuk Batmaz[1], Michel de Mathelin[1] and Birgitta Dresp-Langley[1]*

**Abstract:** In image-guided surgical tasks, the precision and timing of hand movements depend on the effectiveness of visual cues relative to specific target areas in the surgeon's peri-personal space. Two-dimensional (2D) image views of real-world movements are known to negatively affect both constrained (with tool) and unconstrained (no tool) hand movements compared with direct action viewing. Task conditions where virtual 3D would generate and advantage for surgical eye-hand coordination are unclear. Here, we compared effects of 2D and 3D image views on the precision and timing of surgical hand movement trajectories in a simulator environment. Eight novices had to pick and place a small cube on target areas across different trajectory segments in the surgeon's peri-personal space, with the dominant hand, with and without a tool, under conditions of: (1) direct, (2) 2D fisheye camera and (3) virtual 3D viewing (head-mounted). Significant effects of the location of trajectories in the surgeon's peri-personal space on movement times and precision were found. Subjects were faster and more precise across specific target locations, depending on the viewing modality. The significant interactions between

## ABOUT THE AUTHORS

Anil Ufuk Batmaz is a graduate student at the University of Strasbourg and holds Master Degrees in Electrical/Electronic Engineering and Computational Engineering from TOBB University of Economics and Technology in Ankara.

Michel de Mathelin holds a PhD in Engineering and Computer Science from Carnegie Mellon University in Pittsburgh. He is a Senior Fellow of the *IEEE*, Professor of Robotics at the University of Strasbourg, and also Vice President of the University.

Birgitta Dresp-Langley, the principal investigator of this project, holds a PhD in Cognitive Psychology from Paris Descartes University and is a Senior Research Scientist with the CNRS. Our research addresses fundamental issues relative to human performance optimization in simulator training for image-guided surgery and is carried out in close collaboration with the IRCAD, the Hôpitaux Universitaires de Strasbourg (HUS) and the Robotics Department of the ICube Laboratory, of which Michel is also the Director.

## PUBLIC INTEREST STATEMENT

The article addresses critical issues relative to human performance optimization for interventional safety in laparoscopic surgery and robot-assisted surgery. We show results from psychophysical experiments on novice surgeons in our surgical simulator environment *EXCALIBUR*, specifically designed for studying the speed and the precision of hand movements and hand-tool operations under conditions of 2D and 3D image guidance. The data lead to conclude that the relative effectiveness of a 2D or 3D image view for precise and swift hand-tool operations is dependent on the direction of surgical hand movements, and on the spatial position of target areas in the surgeon's peri-personal, or interventional space. A major advantage of 3D image viewing compared with 2D image viewing was not found.

viewing conditions and trajectory locations in peri-personal space explain why 3D viewing is not necessarily an advantage to surgical precision.

**Subjects: Bioscience; Medical Devices; Surgery**

## 1. Introduction

Image-guided hand-tool movements, as in laparoscopic surgery, constrain the surgeon to process critical information about what his/her hands are doing in a real-world environment while looking at a two-dimensional (2D) or three-dimensional (3D) representation of that environment displayed on a monitor. This represents a challenge for the surgeon compared with the natural direct scene viewing in traditional interventions because it involves complex perceptual and cognitive transformations for ensuring optimally precise and safe interventions (e.g. Derossis, Antoniuk, & Fried, 1999; Wentink, 2001). Veridical information about real-world depth is missing from the image representations, and the surgeon is looking sideways or straight ahead at a monitor instead of looking down directly at his hands in the scene of intervention. This lack of direct visual feed-back incurs measurable costs in terms of reduced comfort during task execution, longer times of intervention, or lesser precision, as previous research has clearly shown (Batmaz, de Mathelin, & Dresp-Langley, 2016a, 2016b, 2017; Gallagher, Ritter, Lederman, McClusky, & Smith, 2005; Huber, Taffinder, Russell, & Darzi, 2003; Wilson et al., 2011). Whether virtual 3D image viewing could help overcome these drawbacks has remained a controversial issue. While some have found that 3D viewing significantly improves surgical task performance in both novices and experts (Bhayani & Andriole, 2005; Bueß, van Bergen, Kunert, & Schurr, 1996; Sakata, Grove, Hill, Watson, & Stevenson, 2017; Storz, Buess, Kunert, & Kirschniak, 2012; Taffinder, Smith, Huber, Russell, & Darzi, 1999; Tanagho et al., 2012; Votanopoulos, Brunicardi, Thornby, & Bellows, 2008), others have found equivalent or worse performance with 3D viewing compared with natural 3D or 2D screen views (Chan et al., 1997; Hanna, Shimi, & Cuschieri, 1998; Jones, Brewer, & Soper, 1996; Mueller, Camartin, Dreher, & Hänggi, 1999). Differences in task demands, complexity, and inherent affordance levels (Blavier & Nyssen, 2014) as well as inter-individual differences in adaptive goal-setting strategies of novices (Batmaz et al., 2016b, 2017) could account for discrepancies in results from similar 3D viewing systems. In a recent study by Sakata et al. (2017), it was found that a surgical camera system moving along with the surgical tool in stereoscopic viewing mode produced better depth judgments and faster task execution, in novice and expert surgeons, compared with the 2D viewing mode. To test whether stereoscopic viewing through a headset moving along with the head of the surgeon would provide a performance advantage to 2D screen views, we (Batmaz et al., 2017) used head-mounted stereoscopic viewing (Oculus DK2) in our simulator environment *EXCALIBUR* (Batmaz et al., 2016a, 2016b, 2017). Head-mounted stereoscopic task viewing provides the surgeon with a 3D view of the task scene, and helps overcome previously identified problems relative to viewing position (e.g. Det, Mijerink, Hoff, Totté, & Pierie, 2009; Haveran et al., 2007) by significantly reducing muscle fatigue during interventions (Maithel, Villegas, Stylopoulos, Dawson, & Jones, 2005). In our study, we found no advantage of stereoscopic 3D viewing through the head-mounted device compared with 2D views from an optimally positioned 2D screen. Performance scores in terms of execution time and positional accuracy (on–target precision) were superior in the 2D screen viewing conditions, with equivalent scores for 2D fisheye camera views and corrected 2D views. All subjects were novices with above average spatial ability, as is required for surgical practice.

In surgery, what matters most is precision. In our previous work (Batmaz et al., 2016a, 2016b, 2017) we had focussed on studying the effects of different viewing modalities on positional accuracy, i.e. the precision with which an object was placed on a reference target in the real-world action field (RAF) of our simulator environment *EXCALIBUR*. Effects of target location on the RAF were not studied. Here, we are investigating how 2D and 3D viewing systems affect the precision of the surgeon's hand movements in terms of goal-directed reaching operations across different segments of the RAF, with nearer and further away target areas in the surgeon's peri-personal space. Reaching operations in peri-personal space are encoded topologically in terms of step-by-step representations of hand or hand-tool trajectories (e.g. Davare, Zénon, Desmurget, & Olivier, 2015; Schall, Stuphorn, & Brown, 2002). Movements constrained by the use of a tool (Desmurget, Jordan, Prablanc, & Jeannerod, 1997) and unconstrained movements of the bare hand (e.g. Desmurget et al., 1997) produce different trajectory shapes, with greater or lesser angles of curvature, and are affected to a greater or lesser extent by the eccentricity of targets in the action field (e.g. Desmurget et al., 1997; Van der Graaff, Brenner, & Smeets, 2016). Moreover, reaching movements executed with a tool are known to extend visual-tactile perceptual interaction for eye-hand coordination to locations further away in peri-personal space compared with unconstrained reaching (Farnè & Làdavas, 2000; Longo & Lourenco, 2006; Maravita, 2001; Maravita & Iriki, 2004;). It can therefore be expected that the effects of 2D and 3D viewing modes on constrained vs. unconstrained hand movements will not be the same across target locations in the surgeon's peri-personal space, which is the major working hypothesis motivating the present study.

To test for the predicted interactions between viewing mode, type of surgical hand movement (with or without tool) and target location, we recorded the timing and the spatial coordinates of hand movements of novices picking up a small foam object from a departure point and placing it on five successive target areas at nearer and further away locations on the RAF of our simulator environment *EXCALIBUR*. Subjects had to use their dominant hand, and performed the task with and without a tool under conditions of direct top-down "natural" viewing through a glass pane, 2D screen viewing, and stereoscopic 3D viewing through a head-mounted device. The order of task conditions was rigorously counterbalanced in the experimental design.

## 2. Material and methods
Our study uses the computer controlled perception-action platform *EXCALIBUR* (Batmaz et al., 2016a, 2016b, 2017), designed for an image-based analysis of surgical task parameters relative to time and precision in multi-step pick-and-place tasks, as explained in greater detail here below.

### 2.1. Ethics
The study was conducted in conformity with the Helsinki Declaration relative to scientific experiments on human individuals, with the full approval of the ethics board of the corresponding author's host institution (CNRS). All participants were volunteers and had provided written informed consent.

### 2.2. Subjects
Four healthy right-handed men ranging in age between 20 and 45 and four healthy right-handed women ranging in age between 20 and 45 participated in this study. They were all professionals in administrative careers, with normal or corrected-to normal vision, and naive to the scientific hypotheses underlying the experiments. Pre-screening interviews were conducted to make sure that none of the selected participants had any particular experience in tool-mediated mechanical or surgical procedures. Participants' handedness was assessed using the Edinburgh inventory for handedness designed by Oldfield (1971) to confirm that they were all true right-handers. They were screened for spatial ability on the basis of the Perspective Taking Spatial Orientation Test (PTSOT) developed by Hegarty and Waller (2004), which permits evaluating the ability of individuals to form three-dimensional mental representations of objects and their relative localization and orientation on the basis of merely topological (i.e. non axonometric) visual data displayed two-dimensionally on a sheet of paper or a computer screen. All participants scored successful on 10 or more of the 12 items of the test, which corresponds to spatial ability above average, as would be required for surgery.

### 2.3. Experimental platform: hardware and software
The experimental platform is a combination of hardware and software components designed to test the effectiveness of varying visual environments for image-guided action in real world (Figure 1). The main body of the device contains adjustable horizontal and vertical aluminum bars connected to a stable but adjustable wheel-driven sub platform. The main body can be resized along two different

**Figure 1. Experimental setup for the different viewing conditions: (a) direct (natural 3D), (b) 2D fisheye camera and (c) head mounted virtual 3D stereoscopic.**

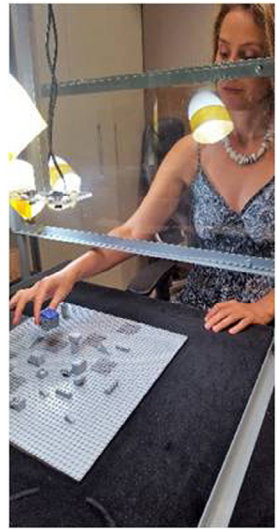

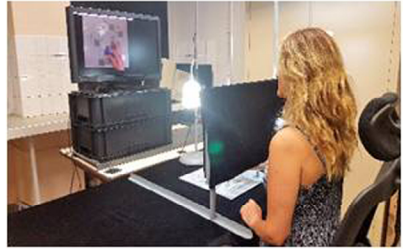

**(b)**

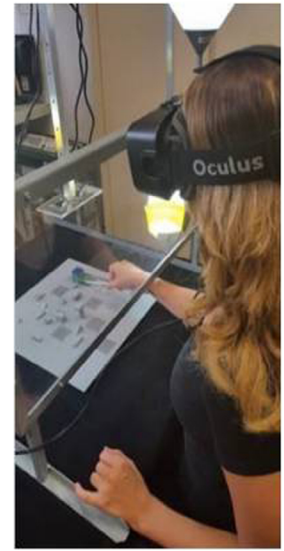

**(a)**

**(c)**

axes in height and in width, and has two HD USB cameras (ELP, Fisheye Lens) fitted into the structure for monitoring the real-world action field from a stable vertical height, which was 60 cm here in this experiment. In this study here, the single 2D camera view was generated through one of the two 120° fisheye lens cameras, both fully adjustable in 360°, connected to a small piece of PVC. For stereoscopic 3D viewing, two camera views for left and right were generated. The video input received from the cameras was processed by a DELL Precision T5810 model computer equipped with an Intel Xeon CPU E5-1620 with 16 Giga bytes memory (RAM) capacity at 16 bits and an NVidia GForce GTX980 graphics card. Experiments were programmed in Python 2.7 using the Open CV computer vision software library. The computer was connected to a high-resolution color monitor (EIZO LCD "Color Edge CG275 W") with an inbuilt color calibration device (colorimeter), which uses the Color Navigator 5.4.5 interface for Windows. The colors of objects visualized on the screen were matched to RGB color space, fully compatible with Photoshop 11 and similar software tools. The color coordinates for RGB triples were retrieved from a look-up table. The screen luminance values for calculating the object contrasts displayed for an experiment are given by the output of the EIZO auto-calibration procedure in candela per square meter ($cd/m^2$). All values were cross-checked with standard photometry using an external photometer (Cambridge Research Instruments) and interface software.

## 2.4. Object movements in the real-world action field

The Real-world Action Field (as of now referred to as the RAF) consisted of a classic square shaped (45 × 45 cm) light grey LEGO© board available worldwide in the toy sections of large department stores. Six square-shaped (4.5 × 4.5 cm) target areas were painted on the board at various locations in a medium grey tint (acrylic). In-between these target areas, small LEGO© pieces of varying shapes and heights were placed to add a certain level of 3D complexity to both the visual configuration and the task. The object that had to be placed on the target areas in a specific order was a small (3 × 3 × 3 cm) cube made of light plastic foam resistant to deformation. Five sides of the cube were painted in the same medium grey tint (acrylic) as the target areas. One side, which was always pointing upwards in the task, was given an ultramarine blue tint (acrylic) to permit tracking object positions. A medium sized barbecue tong with straight ends was used for manipulating the object in the conditions "with tool". The tool-tips were given a matte fluorescent green tint (acrylic) to permit tool-tip tracking.

cogent ••medicine

### 2.5. Object movements displayed on 2D monitor

The video input received by the computer from the HD USB camera generated the raw image data. One single pixel on the screen was 0.32 mm. Real-world data and visual display data were scaled psychophysically for each observer, i.e. the image size was adjusted for each subject to ensure that the visual display matched as closely as possible the scale of the RAF and target areas seen by the individual in the real world. The luminance (L) 289 of the light gray RAF visualized on the screen was 33.8 cd/m$^2$ and the luminance of the medium gray target areas was 15.4 cd/m$^2$, producing a target/background contrast (Weber contrast: ((Lforeground − Lbackground)/Lbackground)) of −0.54. The luminance of the blue ($x = 0.15$, $y = 0.05$, $z = 0.80$ in CIE color space) object surface visualized on the screen was 3.44 cd/m$^2$, producing Weber contrasts of −0.90 with regard to the RAF, and −0.78 with regard to the target areas. The luminance (29.9 cd/m$^2$) of the green ($x = 0.20$, $y = 0.70$, $z = 0.10$ in CIE color space) tool-tips produced Weber contrasts of −0.11 with regard to the RAF, and 0.94 with regard to the target areas. All luminance values for calculating the object contrasts visualized on the screen were obtained on the basis of standard photometry using an external photometer (Cambridge Research Instruments) and interface software.

### 2.6. Object movements displayed in 3D through head-mounted device

The video input received by the computer from the HD USB cameras were fed into a computer vision software (written in Python 2.7 for Windows) which transforms the video data into a stereoscopic 3D image, displayed on the head-mounted screen of the OCULUS DK2 (www.oculus.com/dk2). Real-world data and visual display data were scaled psychophysically for each observer, i.e. the image size was adjusted for each subject to ensure that the visual display subjectively matched the scale of the RAF seen in the real world as closely as possible. In all the image-guided conditions (2D and 3D), the temporal matching of visual display and real-world data was ensured by a computer algorithm which communicates with the internal clock of the CPU.

### 2.7. Experimental procedure

The experiments were run under conditions of free viewing, with illumination levels that can be assimilated to daylight conditions. The RAF was illuminated by two lamps (40 Watt, 6,500 K) which were constantly lit during the whole duration of an experiment. Participants were comfortably seated at a distance of approximately 75 cm from the RAF in the direct viewing condition (Figure 1(a)). The monitor was placed straight ahead of the individuals, and there was no lateral offset from the forearm motor axis. The screen was about 150 cm away from their eyes (Figure 1(b)). To compensate for the change in image size on the screen with the change in body-to-screen distance, the image on the screen was adjusted, ensuring that the perceived scale of the RAF displayed in an image subjectively matched the perceived scale of the RAF when viewed directly. Seats were adjusted individually in height at the beginning of a session to ensure that the image displayed on the monitor was slightly higher than the individual's eyes when looking straight at the screen, which is a near-optimal position given that the optimal monitor height is deemed to be one slightly lower than the eye-level. All participants were given a printout of the targets-on-RAF configuration with white straight lines indicating the ideal object trajectory, and the ordered (red numbers) target positions the small blue cube object had to be placed on in a given trial set of the positioning task (Figure 2), always starting from zero, then going to one, to two, to three, to four, to five, and back to position zero. Participants were informed that they would have to position the cube with their dominant hand "as precisely as possible on the center of each target, as swiftly as possible, and in the right order, as indicated on the printout". They were also informed that they were going to be asked to perform this task under different conditions of object manipulation: with their bare right hand or using a tool, while viewing the RAF and their own hand directly in front of them, on a computer screen as a 2D image, or through the head-mounted Oculus device showing them a 3D image. All participants grasped the object with the thumb and the index of their right hand from the right-hand side in the bare-handed manipulation condition, and from the front with tongs held in their right hand when the tool was used. Before starting the first trial set, the participant could look at the printout shown in Figure 2 for as long as he/she wanted. When they felt confident that they remembered the target order well enough to do the task, the printout was taken away from them and the experiment was started. In the direct wiewing condition, participants saw the RAF and what their hands were doing in top down view through a glass window (Figure 1(a)). In the other viewing conditions, the subjects had to look at a top-down 2D fisheye camera view (Figure 1(b)) or a 3D virtual image view of the RAF seen through a head-mounted device (Figure 1(c)). The order of the different conditions of manipulation and visual feed-back was counterbalanced between subjects and sessions, of which there were two per condition and subject.

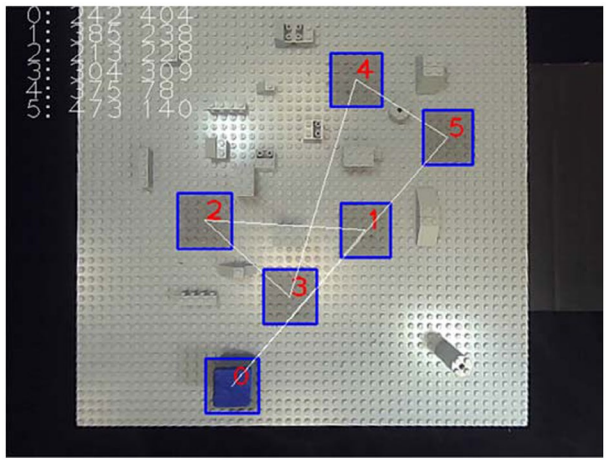

**Figure 2. Screenshot view of the RAF, with the "ideal" reference trajectory, indicated here by the white line. The red numbers show the successive target locations on which the object was to be placed, from the starting point zero to positions one, two, three, four, five, and then back to zero. This order yields six reference trajectory segments, named s1 for movements from zero to one, s2 for movements from one to two, s3 for movments from two to three, s4 for movements from three to four, s5 for movements from four to five and s6 for movements from five back to zero. Segments s2 and s5 have small 3D obstacles on the "ideal" reference line. Segment s3 has a small obstacle to the left of the reference line. Segments s1, s4 and s6 are obstacle-free. The number in the upper left show *x* and *y* data for the five target centers (screen coordinates in pixels). This image was given to subjects at the beginning of the familiarization trials for memorizing the order in which the object had to be placed on the different targets, as swiftly as possible and without making mistakes.**

### 2.8. Familiarization trials and data coding

Data from fully completed trial sets only were recorded. A fully complete trial set consists of a set of positioning operations, starting from location zero to one (trajectory segment s1), from one to two (trajectory segment s2), from two to three (trajectory segment s3), from three to four (trajectory segment s4), from four to five (trajectory segment s5) and then back to starting point zero (trajectory segment s6), performed without dropping the object, and without errors in the positioning order. Whenever such occurred (this happened only incidentally at the beginning during the familiarization trials), the trial set was aborted immediately, and the participant started from scratch in that specific condition. As stated above, ten fully completed trial sets were recorded for each combination of factor levels. The whole trajectory of anticipated object displacements being sub-divided into six segments, as explained above, the blue foam-cube's top center position was followed and recorded for each of these segments. The software started collecting these positional data from the moment the blue object was picked up et location zero. How the ideal object location on a target was computed from the camera views is shown here in Figure 3. The video frame rate was between 25 and 30 Hz (less than 40 ms in the temporal domain). The standard error of video trajectory estimations was less than two pixels. The trajectory data, counted in pixels, were saved to a cloud file.

*cogent* · medicine

**Figure 3. Schematic illustration explaining how the computer vision algorithm codes the ideal object-on-target position from the video-images. This provides a measure of positional accuracy (in pixels), exploited in our previous study (Batmaz et al., 2017). In this study here, we focus on from-target-to-target movement precision (in pixels) where the shortest distance between targets, indicated by the white lines in Figure 2, serves as reference trajectory.**

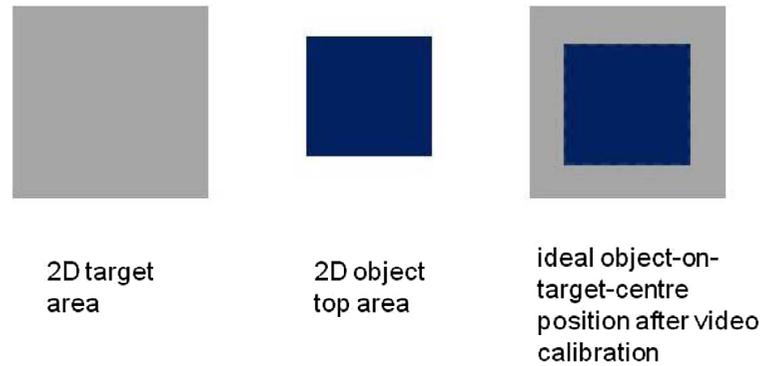

2D target area

2D object top area

ideal object-on-target-centre position after video calibration

## 3. Results

In a first step, we performed point-by-point discrete sampling of spatial coordinates ($x$, $y$) of individual object movement trajectories in the different experimental conditions. These were then plotted against the ideal reference trajectory line (cf. Figure 2) to graphically represent real-world object movement trajectory offsets in terms of planar ($x$, $y$) sketches (Figure 4). The average lateral offset per subject (averaged over the ten repeated trials per condition), experimental condition, trajectory segment and session was computed as an indicator of the precision of the hand-object movements. The average time per subject (averaged over the ten repeated trials per condition) between pick-and-place operations for each trajectory segment, experimental condition and session was computed as an indicator of hand movement times.

### 3.1. Qualitative analysis of the shape of movements

The sampled real-world object trajectories across the different segments of the RAF are indicated by the blue points in the graphs in Figure 4, which shows trajectory data for constrained and unconstrained object movements in the different visual-feed-back conditions. The blue trajectory points in the graphs on the left of Figure 4 exhibit the expected curved shapes of unconstrained hand movements, when no tool is used to displace the object. Figure 4 also shows that constrained movements with use of a tool (graphs on right) do not necessarily follow the expected straight line path. The variations in shape of the sampled real-world trajectories for constrained and unconstrained movements in the surgeon's peri-personal space suggest complex effects of the type of visual feed-back given, type of object movement to be realized, and target position or eccentricity on the RAF, which is consistent with results from previous work on constrained and unconstrained hand movements, introduced here above. The results also point toward critical interactions between viewing conditions and target-to-target trajectory positions. Five-way Analysis of Variance (ANOVA) was performed on the average data for precision (measured in pixels) and time of target-to-target hand movements (measured in milliseconds) to check where significant interactions occur in the data.

### 3.2. Four-way ANOVA on individual means for precision and movement time

For each $x$, $y$ coordinate sampled, we calculated its lateral offset from the ideal reference trajectory, which is indicated by the straight green lines in the graphs in Figure 4. The average individual data from each experimental condition were committed to four-way analysis of variance (ANOVA) on the basis of a Cartesian design plan $P_8 \times V_3 \times M_2 \times T_6 \times S_2$, with eight participants ($P_8$), three levels of viewing ($V_3$) with direct real *vs* 2D fisheye *vs* virtual 3D, two levels of movement ($M_2$) with constrained *vs* unconstrained, six levels of the trajectory segment location factor ($T_6$), and two levels of the session factor ($S_2$). With this design plan, we have a total of 576 means for the dependent variable lateral

**cogent** ∘∘medicine

**Figure 4. Real-world object movement trajectories, indicated by the blue points here, in the different visual feed-back conditions. *Real 3D* direct viewing, without tool use (a) and with tool use (b), produces markedly smaller lateral offsets of real world trajectory points from the ideal reference trajectory (indicated by the straight green lines) compared with *stereoscopic virtual 3D*, without tool use (c) and with tool use (d), or *2D fisheye* camera views (e), (f). Unconstrained 3D and 2D image guided movements (c and e) across the different trajectory segments produce more scatter compared with the constrained (d and f) movements.**

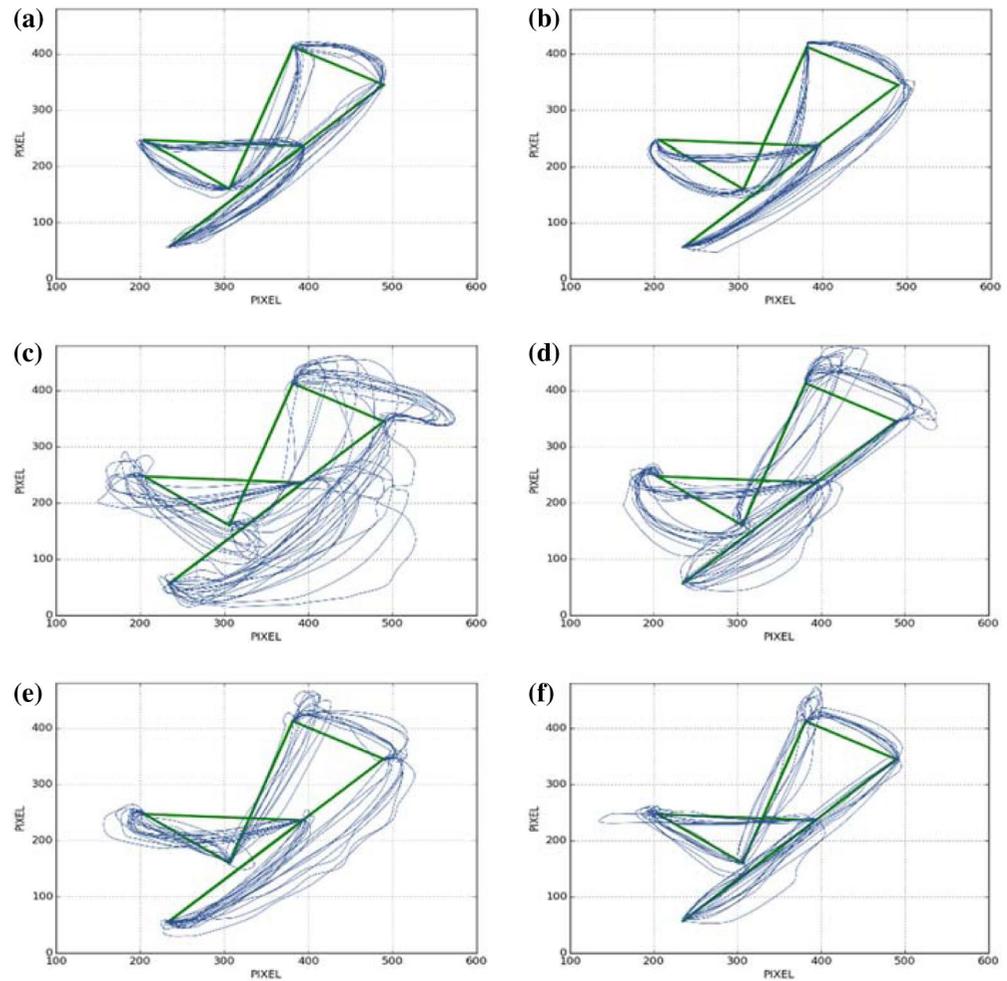

**Table 1. Means relative to time and precision, with standard errors (SEM), F statistics and probability limits (*p*) from the five-way ANOVA**

| Factor | Level | Movement time (milliseconds) | | | Ideal-to-real trajectory offset (pixels) | | |
|---|---|---|---|---|---|---|---|
| | | *M* | SEM | *F*-value | *M* | SEM | *F*-value |
| Viewing | Direct | 1,010 | 20 | $F_{(2,575)} = 240.13$; $p < 0.001$ | 13 | 0.2 | $F_{(2,575)} = 275.84$; $p < 0.001$ |
| | 2D | 1,770 | 30 | | 15 | 0.4 | |
| | Virtual 3D | 1,900 | 60 | | 33 | 0.8 | |
| Manipulation | No tool | 1,460 | 20 | $F_{(1,575)} = 29.15$; $p < 0.001$ | 19 | 0.5 | NS |
| | Tool | 1,650 | 40 | | 18 | 0.6 | |
| Session | Session 1 | 1,688 | 25 | $F_{(1,575)} = 19.81$; $p < 0.001$ | 19 | 0.3 | NS |
| | Session 2 | 1,477 | 20 | | 17 | 0.3 | |
| Trajectory location | X0 | 1,673 | 40 | $F_{(5,575)} = 45.42$; $p < 0.001$ | 16 | 0.5 | $F_{(5,575)} = 79.22$; $p < 0.001$ |
| | X1 | 973 | 38 | | 30 | 0.7 | |
| | X2 | 1,545 | 40 | | 14 | 0.3 | |
| | X3 | 1,620 | 42 | | 15 | 0.6 | |
| | X4 | 1,785 | 43 | | 19 | 0.6 | |
| | X5 | 1,749 | 50 | | 17 | 0.6 | |

trajectory offset (precision), and 576 means for the dependent variable movement time. Table 1 shows the means, standard errors (SEM), *F*-statistics and probability limits (*p*) from this ANOVA.

### 3.3. Effects of viewing

Significant effects of viewing modality on movement times and movement precision were found. Subjects were fastest and the most precise under conditions of direct (real 3D) viewing, where the shortest movement times (Table 1, left) and the smallest lateral offsets from the ideal movement trajectory (Table 1, right) were observed. Post-hoc paired comparisons using the Holm-Sidak method were performed to check which of the differences between the three levels of the viewing factor were significant. For *movement time*, the difference in means between 2D and *virtual* 3D (*d* = 130 ms) signaled non-significant. The differences in means between *direct* and 2D (*d* = 660 ms) and between *direct* and *virtual* 3D (*d* = 890 ms) signaled significant with $t(1,1)$ = 11.37, $p < 0.001$ and $t(1,1)$ = 13.18, $p < 0.001$ respectively. For *precision*, the difference in means between *direct* and 2D (*d* = 2 pix) signaled non-significant. The differences in means between 2D and 3D (*d* = 18 pix) and between *direct* and *virtual* 3D (*d* = 20 pix) signaled significant with $t(1,1)$ = 13.74, $p < 0.001$ and $t(1,1)$ = 15.33, $p < 0.001$ respectively. ANOVA signaled statistically *significant interactions* between the viewing factor and the trajectory location factor on hand movement time ($F(10, 575)$ = 4.33, $p < 0.01$), and on precision ($F(10, 575)$ = 14.46, $p < 0.001$). These interactions are adressed in further detail here below.

### 3.4. Effects of type of hand movement

A significant effect of the manipulation factor or type of hand movement on *movement time* was found, where subjects were significantly faster (Table 1, left) when no tool was used (unconstrained hand movements). The effect of the manipulation factor on precision signaled non-significant (Table 1, right). Significant interactions between type of hand movement and any of the other factors were not found.

### 3.5. Effects of session

A significant effect of the session factor on *movement time* was found, where subjects were significantly faster (Table 1, left) in the first session. The effect of session on precision signaled non-significant (Table 1, right). Significant interactions between the session factor and any of the other factors were not found.

### 3.6. Effects of trajectory segment location as a function of viewing modality

Significant effects of the location of trajectories in the surgeon's peri-personal space (within the limits of the RAF) on movement times (Table 1, left) and movement precision (Table 1, right) were found. Subjects were faster and more precise across specific locations, depending on the viewing modality as signaled by the significant interactions between the viewing and trajectory location factors, with probability limits (*p* < 0.001 for time and precision), as stated above. To find out which differences in mean scores for time and precision between trajectory locations were significant for a given level of the viewing factor, *post hoc* paired comparisons were performed using the Holm-Sidak method. Comparisons that signaled a statistically significant difference are summarized in Table 2 for each of the three levels of the viewing factor, with effect sizes, t-statistics, and the corresponding probability limits.

The viewing modality producing the largest number of significant differences in hand movement times and precision across trajectory segment locations is the virtual 3D condition (see Table 2). The 2D and the direct viewing conditions produced fewer significant differences between locations. More importantly, the trajectory segment locations producing significant differences in hand movement

**✴ cogent** • medicine

**Table 2. Results of the paired *post hoc* comparisons between levels of the trajectory segment factor for each of the three levels of the viewing factor, with effect sizes and t-statistics**

| Significant interactions | *Post-hoc* comparison | Time (milliseconds) | | Ideal-to-real trajectory offset (pixels) | |
|---|---|---|---|---|---|
| | | *d* | *t* (Holm-Sidak) | *d* | *t* (Holm-Sidak) |
| Viewing × trajectory location on RAF | X1 × X2 in *direct* | 321 | 2.98; *p* < 0.01 | 11 | 6.57; *p* < 0.01 |
| | X1 × X3 in *direct* | 357 | 3.32; *p* < 0.01 | 10 | 5.82; *p* < 0.01 |
| Effects of *trajectory location in direct* | X1 × X5 in *direct* | 462 | 4.30; *p* < 0.01 | 7 | 4.43; *p* < 0.01 |
| | X1 × X0 in *direct* | 452 | 4.20; *p* < 0.01 | 6 | 3.64; *p* < 0.01 |
| | X1 × X4 in *direct* | 447 | 4.16; *p* < 0.01 | 6 | 3.44; *p* < 0.01 |
| | X4 × X2 in *direct* | 126 | NS | 5 | 3.15; *p* < 0.05 |
| | X0 × X2 in *direct* | 130 | NS | 5 | 2.93; *p* < 0.05 |
| Effects of *trajectory location in virtual 3D* | X1 × X2 in *3D* | 513 | 4.77; *p* < 0.01 | 28 | 17.95; *p* < 0.001 |
| | X1 × X0 in *3D* | 1,103 | 10.25; *p* < 0.001 | 9 | 5.75; *p* < 0.01 |
| | X1 × X3 in *3D* | 736 | 6.84; *p* < 0.01 | 25 | 15.30; *p* < 0.001 |
| | X1 × X5 in *3D* | 809 | 7.52; *p* < 0.001 | 21 | 12.91; *p* < 0.001 |
| | X1 × X4 in *3D* | 1,021 | 9.49; *p* < 0.001 | 48 | 19.47; *p* < 0.001 |
| | X4 × X2 in *3D* | 508 | 4.72; *p* < 0.01 | 11 | 6.88; *p* < 0.01 |
| | X4 × X0 in *3D* | 576 | 4.99; *p* < 0.01 | 11 | 6.85; *p* < 0.01 |
| | X4 × X3 in *3D* | 285 | NS | 8 | 4.83; *p* < 0.01 |
| | X5 × X2 in *3D* | 296 | 2.75; *p* < 0.05 | 7 | 4.44; *p* < 0.01 |
| | X5 × X0 in *3D* | 262 | NS | 7 | 4.41; *p* < 0.01 |
| Effects of *trajectory location in 2D* | X1 × X3 in *2D* | 850 | 7.89; *p* < 0.001 | 11 | 6.48; *p* < 0.01 |
| | X1 × X4 in *2D* | 967 | 8.99; *p* < 0.001 | 10 | 5.92; *p* < 0.01 |
| | X1 × X0 in *2D* | 547 | 5.09; *p* < 0.01 | 18 | 12.31; *p* < 0.001 |
| | X1 × X2 in *2D* | 882 | 8.19 *p* < 0.001 | 9 | 5.57; *p* < 0.01 |
| | X1 × X5 in *2D* | 956 | 8.81; *p* < 0.001 | 9 | 5.93; *p* < 0.01 |

times and precision are not same in the different viewing modalities. These results point toward a complex interdependency between visual feed-back conditions, the direction and shape of the hand movements from target to target, bearing in mind that some target-to-target trajectories contained small obstacles, and the spatial position of the targets. To gather further insight into this complexity, we plotted the individual means for precision in the different conditions as function of the trajectory segment location.

These individual data are shown in Figure 5, with average lateral deviations from the ideal movement trajectory as a function of trajectory segment locations on the RAF, in the different experimental conditions and sessions.

The data of the men are shown in 5(a), the data of the women in 5(b). The data curves reveal consistent shapes across subjects, with almost invariably the worst precision scores across the trajectory segments X1 and X4, especially under conditions of virtual 3D viewing. Both trajectory segments involve target-to-target movements across a small obstacle in the sideways direction (from left to right) in the peri-personal space of the subject/surgeon.

The longest hand movement times were observed across the trajectory segment X4, which involves the target-to-target hand movement the furthest away in the subject's peri-personal space. This result is shown here in Figure 6, with the average hand movement times in the different conditions plotted as a function of the trajectory segment location.

**Figure 5**. Individual means (here the data were averaged over the ten trials per subject and experimental condition) relative to ideal-to-reference trajectory lateral offsets, in pixels, from session 1 and session 2 plotted as a function of the viewing conditions and the location of the trajectory segments in the surgeon's peri-personal space. Segments are labelled arbitrarily, from "X0" for the first requiring hand movement away from the starting point, to "X5" for the last requiring hand movement back to the starting point. Individual means of the men are shown in 5(a), individual means of the women in 5(b).

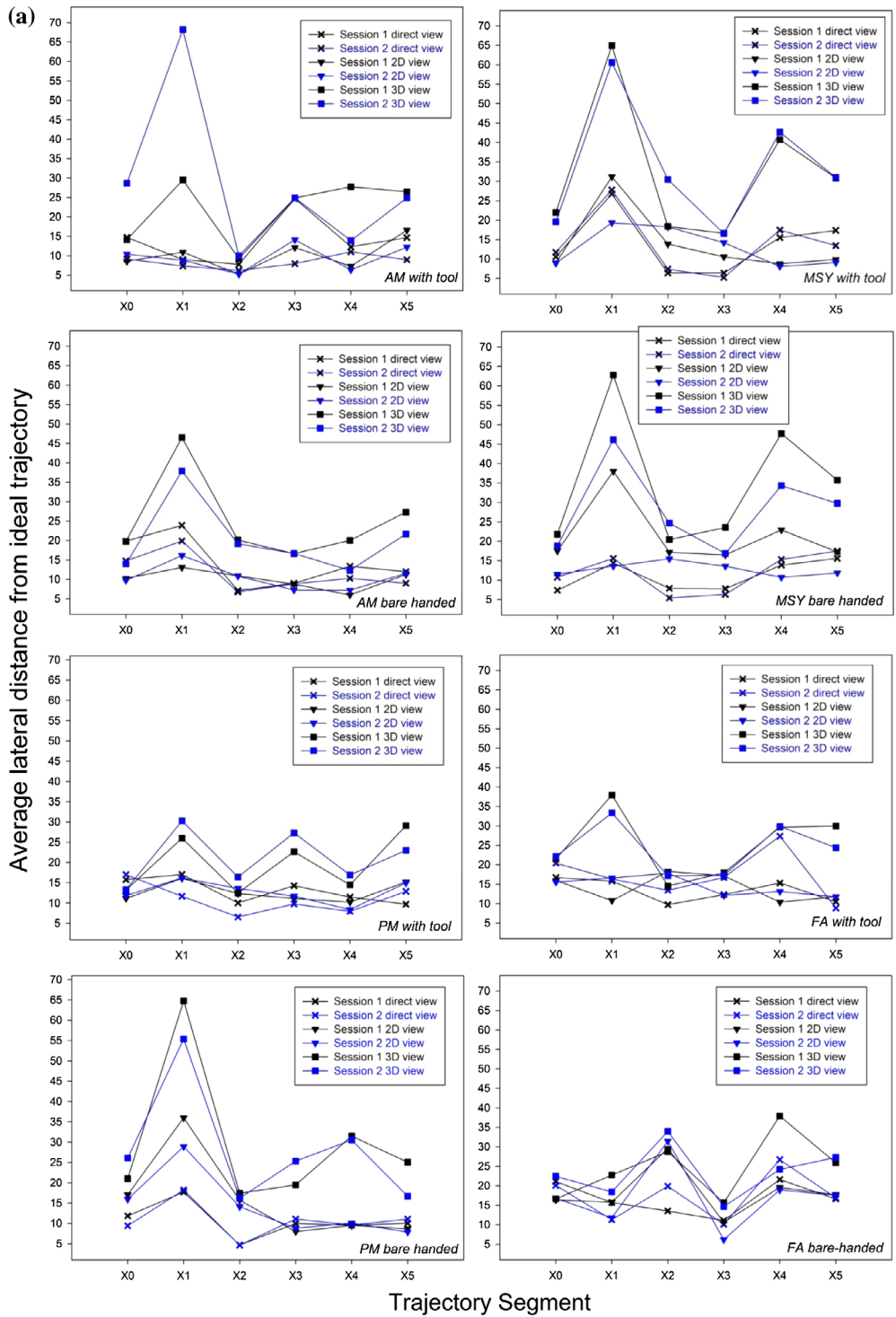

**Figure 5. (Continued).**

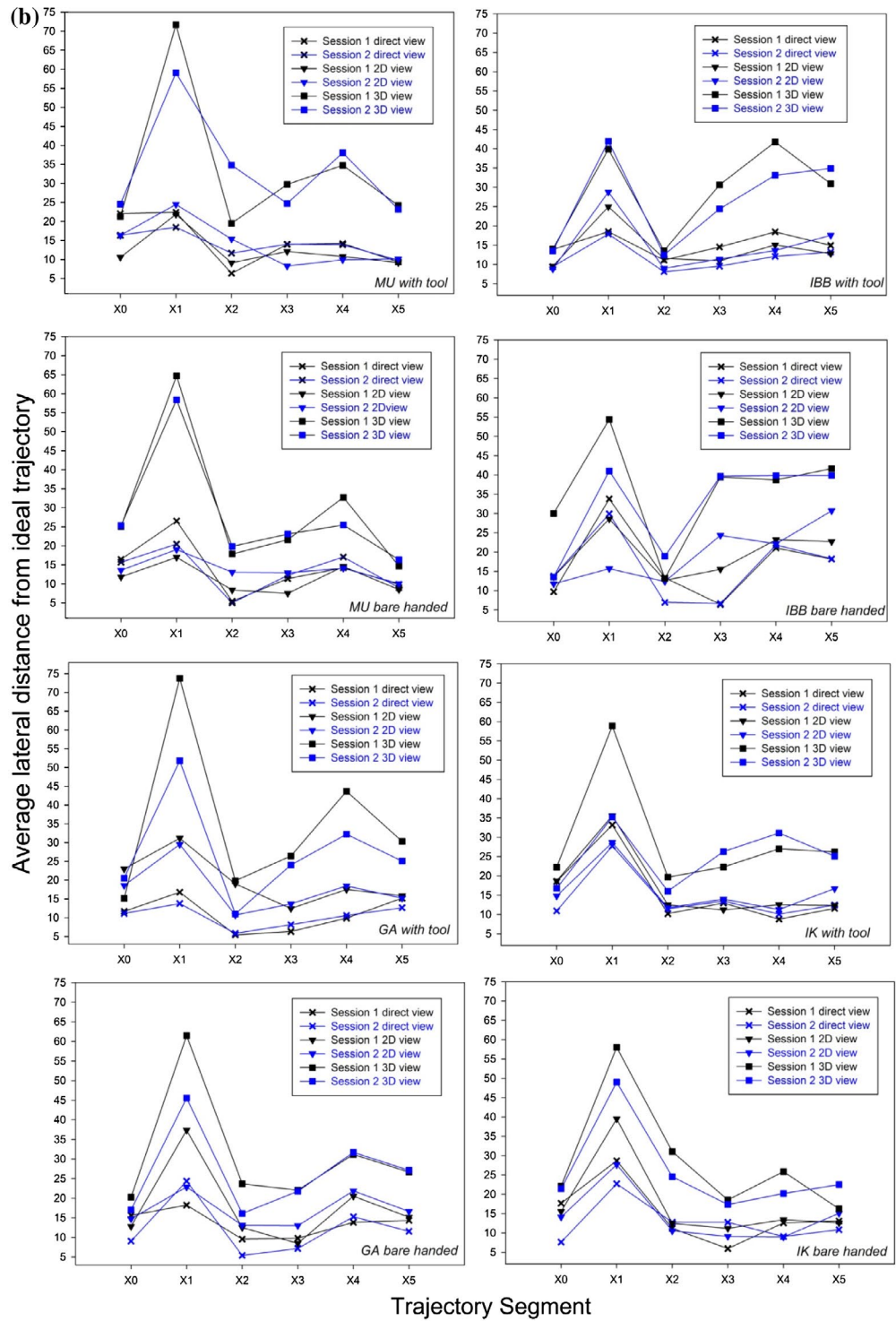

**Figure 6. Group means (here the data were averaged over the ten trials per subject and over subjects) for hand movement time across a trajectory segment, in milliseconds, from session 1 and 2 plotted as a function of the viewing conditions and the location of the trajectory segments in the surgeon's peri-personal space.**

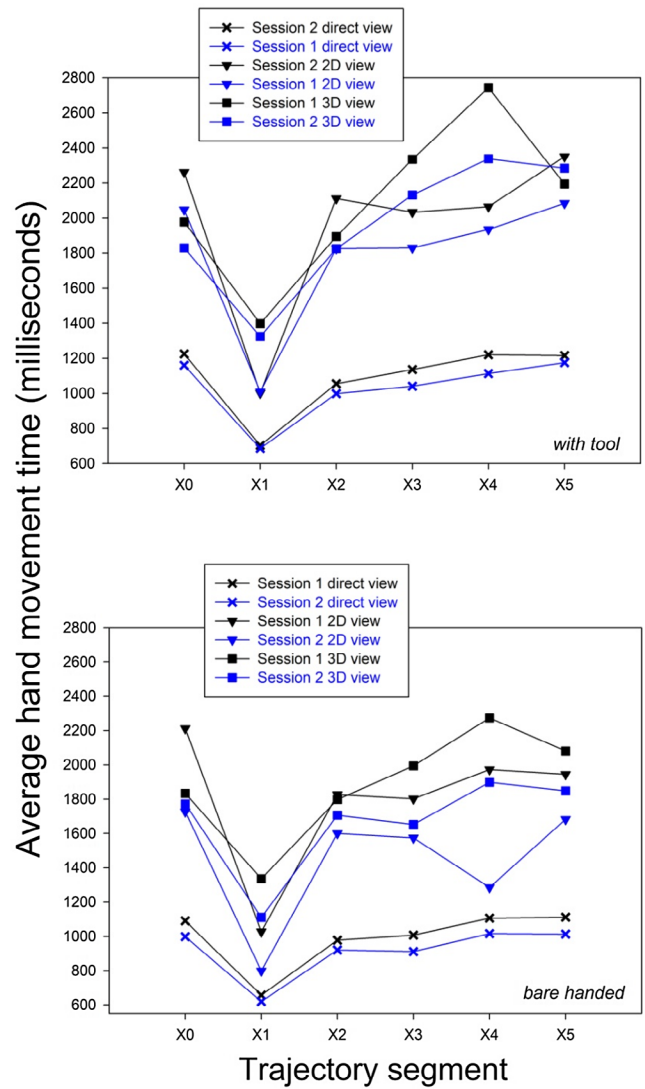

## 4. Discussion

The effects of 2D and 3D viewing modes on the precision and timing of surgical hand movements, whether constrained by tool-use or not, are not the same across target locations in the surgeon's peri-personal space, as clearly shown by the results of our simulator study here. Hand movements away from or back to the body in a pick-and-place simulator task appear less affected by image guidance compared with direct action viewing. Hand movements in a sideways direction (here, to the left) in the surgeon's peri-personal space are markedly less precise with image guidance, especially with stereoscopic 3D viewing. Moreover, at target locations further away in peri-personal space, sideways movements are considerably slowed down with image guidance, especially under conditions of 3D stereoscopic viewing, as shown here. These results suggest that goal-directed hand movements, whether constrained or not, are affected by both the direction of movement and the spatial position (or eccentricity) of target locations in peri-personal space (e.g. Desmurget et al., 1997; Sarlegna et al., 2003). Small obstacles on the target-to-target trajectories affect precision and timing of hand movements, as shown here, and this drawback is not compensated for by a 3D image view, supposed to convey three-dimensional cues to the surgeon to a greater extent than a 2D image view.

The complex interactions between viewing conditions and target location found in our study here point towards possible reasons why some have found that 3D viewing significantly improves surgical task performance in both novices and experts (Bhayani & Andriole, 2005; Bueß et al., 1996; Sakata et al., 2017; Storz et al., 2012; Taffinder et al., 1999; Tanagho et al., 2012; Votanopoulos et al., 2008), while others observed equivalent or worse performance with 3D viewing compared with natural (real 3D) or 2D screen views (Batmaz et al., 2017; Chan et al., 1997; Hanna et al., 1998; Jones et al., 1996; Mueller et al., 1999). 3D stereoscopic viewing does not help to compensate for the effects of variations in hand movement direction and position in peri-personal space, which are known to affect the control of human arm movements (Haggard & Richardson, 1996; Krakauer & Mazzoni, 2011). Neuropsychological evidence suggests that action influences spatial perception (Humphreys, Riddoch, Forti, & Ackroyd, 2004); the latter, during hand movements in particular, is known to determine an individual's sense of agency, or feeling in control (Balslev, Cole, & Miall, 2007). It can be assumed that the type of tool used for action, and the way in which a camera system captures the movements seen on the screen would have a critical impact on both. When the surgical camera system is part of the surgical tool itself and moves along with the tool in peri-personal space as in a recent study by Sakata et al. (2017), a positive effect of stereoscopic viewing on surgical task execution times was, indeed, found. A state-of-the-art endoscopic 2D/3D camera system (EndoEye Flexlens) built into the tip of the surgical tool was used in that study.

Finally, the subjects in our study here were all novices, with the above-average spatial abilities necessary for surgery, but without any training in image-guided procedures. Such beginners are bound to have more heterogeneous general training backgrounds than expert surgeons. They still need to get used to the image views, whether 2D or 3D, when monitoring their hands moving across peri-personal space (Masia, Casadio, Sandini, & Morasso, 2009). Effective eye-hand coordination under image guidance can only be considered near-optimal once it produces performance scores with stable speed-precision trade-offs in one and the same individual (Batmaz et al., 2017). Getting there involves complex processes of perceptual learning for motor control and action that deserve to be investigated further.

## 5. Conclusions

3D viewing systems do not straightforwardly produce better surgical eye-hand coordination in image guided procedures. The relative effectiveness of 3D technology for the precision and timing of surgical hand movements depends on the type and direction of hand movement required for the intervention, the flexibility of the camera system generating the image views across target locations in the surgeon's peri-personal space, and on the surgeon's individual training level.

**Authors' contributions**

AUB (graduate student) participated in the design of the experimental platform, programmed the software, selected participants, carried out the experiments and ensures data curation. MdM (thesis co-supervisor) contributed ideas to the experimental design and data analysis. BDL (thesis supervisor) designed *EXCALIBUR* and the psychophysical experiments, analyzed data, wrote the manuscript and took care of the revisions in close collaboration with MdM and AUB.

**Funding**

The study was funded by the Initiative D'EXcellence [grant number IDEX 2015-2018] of the University of Strasbourg (Principal Investigator Birgitta DRESP-LANGLEY). Material (computers, screens and hardware) for building *EXCALIBUR* was financed by the CNRS (MI AAP 2015 to Birgitta DRESP). The funding bodies had no role in the study design, data collection, or decision to submit this manuscript for publication.

**Author details**

Anil Ufuk Batmaz[1]
E-mail: aubatmaz@etu.unistra.fr
ORCID ID: http://orcid.org/0000-0001-7948-8093
Michel de Mathelin[1]
E-mail: demathelin@unistra.fr
Birgitta Dresp-Langley[1]
E-mail: birgitta.dresp@unistra.fr
ORCID ID: http://orcid.org/0000-0002-2860-6472
[1] ICube Lab, CNRS and University of Strasbourg, UMR 7357, Strasbourg, France.

❋ cogent •• medicine

## Competing interests

The authors declare no competing interest.

*Cogent Medicine* (ISSN: 2331-205X) is published by Cogent OA, part of Taylor & Francis Group.

**Publishing with Cogent OA ensures:**

• Immediate, universal access to your article on publication

• High visibility and discoverability via the Cogent OA website as well as Taylor & Francis Online

• Download and citation statistics for your article

• Rapid online publication

• Input from, and dialog with, expert editors and editorial boards

• Retention of full copyright of your article

• Guaranteed legacy preservation of your article

• Discounts and waivers for authors in developing regions

**Submit your manuscript to a Cogent OA journal at www.CogentOA.com**

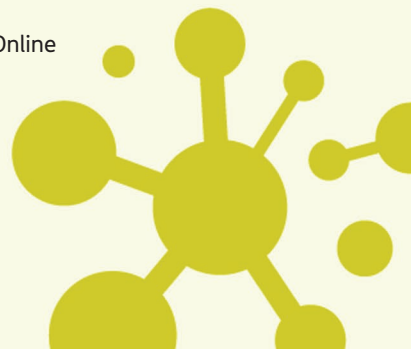